\documentclass[11pt]{article}
\usepackage{amsmath,amssymb,color,epsfig,cite}
\usepackage{graphicx}
\usepackage{setspace}

\usepackage{mathrsfs}

\usepackage{amsfonts}
\newcommand{\hoch}[1]{$\, ^{#1}$}

\makeatletter
\@addtoreset{equation}{section}
\makeatother

\newcommand{\be}{\begin{equation}}
\newcommand{\ee}{\end{equation}}
\newcommand{\bea}{\setlength\arraycolsep{2pt} \begin{eqnarray}}
\newcommand{\eea}{\end{eqnarray}}
\newcommand{\nn}{\nonumber}
\newcommand{\half}{{\textstyle{\frac{1}{2}}}}

\def\ndelta{\delta\hspace{-0.50em}\slash\hspace{-0.05em} }

\def\ft#1#2{{\textstyle{\frac{\scriptstyle #1}{\scriptstyle #2} } }}
\def\fft#1#2{{\frac{#1}{#2}}}

\def\0{{\sst{(0)}}}
\def\1{{\sst{(1)}}}
\def\2{{\sst{(2)}}}
\def\3{{\sst{(3)}}}
\def\4{{\sst{(4)}}}
\def\5{{\sst{(5)}}}
\def\6{{\sst{(6)}}}
\def\7{{\sst{(7)}}}
\def\8{{\sst{(8)}}}
\def\sst#1{{\scriptscriptstyle #1}}

\def\scri{\mathscr{I}}

\def\crampest{\medmuskip = 1mu plus 1mu minus 1mu}
\def\uncramp{\medmuskip = 4mu plus 2mu minus 4mu}

\allowdisplaybreaks

\begin{document}

\begin{flushright}
\hfill{\hfill{MI-TH-1930}}

\end{flushright}

\vspace{15pt}
\begin{center}
{\Large {\bf Dual gravitational charges and soft theorems}}

\vspace{15pt}
{\bf Hadi Godazgar\hoch{1}, Mahdi Godazgar\hoch{2} and 
C.N. Pope\hoch{3,4}}

\vspace{10pt}

\hoch{1} {\it Max-Planck-Institut f\"ur Gravitationsphysik (Albert-Einstein-Institut), \\
M\"uhlenberg 1, D-14476 Potsdam, Germany.}

\vspace{10pt}

\hoch{2} {\it School of Mathematical Sciences,
Queen Mary University of London, \\
Mile End Road, E1 4NS, United Kingdom.}

\vspace{10pt}

\hoch{3} {\it George P. \& Cynthia Woods Mitchell  Institute
for Fundamental Physics and Astronomy,\\
Texas A\&M University, College Station, TX 77843, USA.}

\vspace{10pt}

\hoch{4}{\it DAMTP, Centre for Mathematical Sciences,\\
 Cambridge University, Wilberforce Road, Cambridge CB3 OWA, UK.}

 \vspace{15pt}
 
August 3, 2019

\vspace{20pt}

\underline{ABSTRACT}
\end{center}

\noindent
We consider the consequences of the dual gravitational charges for the 
phase space of radiating modes, and find that they imply a new soft NUT 
theorem. In particular, we argue that the existence of these new charges 
removes the need for imposing boundary conditions at spacelike infinity 
that would otherwise preclude the existence of NUT charges. 
\noindent

\thispagestyle{empty}

\vfill
E-mails: hadi.godazgar@aei.mpg.de, godazgar@phys.ethz.ch, pope@physics.tamu.edu

\pagebreak

\section{Introduction}

Recent investigations \cite{fakenews, dual0, dualex} of the relation between 
Newman-Penrose charges \cite{NP} and the charges corresponding to the 
BMS symmetry group of asymptotically flat spacetimes \cite{BarTro} have 
led to the discovery of two generalisations of BMS charges: 
a generalisation in terms of a subleading $1/r$ expansion and a 
generalisation in terms of a complexification of the BMS charges.  
In this context, the real and imaginary parts of the Newman-Penrose charges 
corresponds to the subleading supertranslation and \emph{dual} 
supertranslation charges at order $1/r^3$.

The dual gravitational charges are given by twisted fields, defined using 
the Levi-Civita on the 2-sphere, and as such do not appear in components 
of the Einstein equation; rather, they are related to NUT charges 
\cite{dual0, porrati}, and may be viewed as the gravitational analogues of 
magnetic monopoles.  As has been emphasised recently in Ref.\ \cite{porrati}, 
and in older works \cite{Ramaswamy,ashtekar}, the existence of 
non-trivial dual charges (at least globally) is intrinsically related to 
the non-trivial topological structure of spacetime, in the same way that 
a gauge connection in electromagnetism can lead to a non-trivial 
magnetic monopole charge.

In this paper, we investigate further the relationship between these new 
dual gravitational charges and the well-known NUT charge \cite{taub,nut} by 
using the Komar integrals for energy and dual charge, which are defined 
for stationary solutions, as guides.   Furthermore, we investigate the 
consequence of the existence of such charges for the gravitational phase 
space at null infinity, which has been discussed in 
Refs.\ \cite{Strom:soft2}, \cite{porrati}, and for the Weinberg soft 
theorems \cite{weinberg}, whose relation to asymptotic symmetries has 
attracted much recent attention \cite{Strom:YM, Strom:soft1, Strom:soft2, Strom:QED, Strominger:2014pwa, Strom:mag, Hawking:2016msc, Sheikh-Jabbari:2016lzm, Hawking:2016sgy, conde, Choi:2017bna, Campoleoni:2018uib, Campiglia:2018dyi, Mao:2018xcw, Himwich:2019qmj, Freidel:2019ohg, Kutluk:2019ghr, Adamo:2019ipt, Alessio:2019cae, Laddha:2019yaj, He:2019ywq, Choi:2019rlz}.  We shall, mainly, 
focus on (dual) supertranslation charges in this paper.

The relation between dual charges and non-trivial topology leads us to 
propose a generalisation of the concept of asymptotic flatness to 
include tensors in a $1/r$ expansion of the metric components that are 
not necessarily regular on the 2-sphere at infinity.  This conclusion is 
reminiscent of the introduction of superrotations as viable BMS 
transformations \cite{Barnich:2009se, BarnichAspects}.  Indeed, it seems 
natural both in terms of the existence of dual BMS charges, and also in 
allowing 
the action of a larger group of asymptotic symmetry generators, that 
such a generalisation be considered.  However, the introduction of fields 
that are 
not necessarily regular on the 2-sphere means that one must be more careful 
in dealing with total derivative terms.  In particular, given that 
previous results on BMS charges \cite{BarTro, fakenews, dual0, dualex} 
have assumed all tensors to be regular, we need to go back and re-evaluate 
the derivations of the charges.  This leads to new 
expressions for the supertranslation charges, with total derivative terms 
that do not necessarily integrate to zero on the 2-sphere.

Following the observations in Ref.\ \cite{dualex}, which found that the usual and dual supertranslation charges are associated with the real and imaginary parts of an appropriate Newman-Penrose scalar, and the strategy employed in Ref.\ \cite{Strom:mag}, we construct a complexified supertranslation charge and investigate its action on the phase space.  We find that the complexified supertranslation charge acts on one mode as a time translation, while it acts on the other mode as a supertranslation.  This is in complete analogy with what happens in 
electromagnetism, as described in Ref.\ \cite{Strom:mag}.  It is worth 
emphasising that we obtain these expected Dirac brackets without the need to 
impose any boundary conditions at spacelike infinity, in contrast to 
Refs.\ \cite{Strom:soft2}, \cite{porrati}.  With hindsight the boundary 
condition required in order to obtain reasonable Dirac brackets in 
Ref.\ \cite{Strom:soft2} is justified, because in that work the dual 
charge has effectively been set to be zero and, for consistency, 
this condition must be reflected in the phase space via the boundary 
conditions imposed thereon.  However, given that one imposes also a 
boundary condition at timelike infinity, it is unsatisfactory from the 
point of view of an initial-value formulation to insist on a boundary condition 
 also at spacelike infinity.  Therefore, we conclude that a satisfactory 
resolution of the phase space problem identified in Ref.\ \cite{Strom:soft2} 
is to include dual supertranslation charges, thereby eliminating the need 
for boundary conditions at spacelike infinity.  Assuming a conservation of 
the complexified charge across spacelike infinity then leads to a 
Weinberg-like soft NUT/graviton theorem, in the same vein as that proposed 
for electromagnetism in Ref.\ \cite{Strom:mag}.

In section \ref{sec:komar}, we compare the usual and dual supertranslation 
charges with Komar integrals, which are defined for stationary spacetimes, 
and we use this comparison to motivate a generalisation of the notion of 
asymptotic flatness to spacetimes with metric components that are 
not necessarily regular on the 2-sphere.  In section \ref{sec:phase}, 
we derive the supertranslation charges associated with these generalised 
asymptotically flat spacetimes, we define a complexified supertranslation 
charge, and we find that its action on phase space is analogous to the 
situation in electromagnetism.  We end in section \ref{sec:soft} by deriving 
a soft NUT/graviton theorem.  We also include two appendices, where we
give detailed constructions of the Kerr and the Taub-NUT metrics in 
Bondi coordinates, up to the first few orders in a $1/r$ expansion. These
examples serve to illustrate the fact that in the case of the Kerr metric,
which has no non-vanishing global dual charges, the Bondi metric coefficients are
completely non-singular on the 2-sphere.  By contrast, when the Taub-NUT 
metric is written in Bondi form, the metric coefficients in the $1/r$ 
are inevitably singular somewhere on the 2-sphere.

\section{Asymptotically flat spacetimes and NUT charges} \label{sec:komar}

Choosing outgoing Bondi coordinates $(u,r,x^I=\{\theta, \phi\})$, we define asymptotically flat spacetimes to be those for which the metric takes 
the form \cite{bondi, sachs}
\begin{equation} \label{AF}
 d s^2 = - F e^{2 \beta} du^2 - 2 e^{2 \beta} du dr + 
r^2 h_{IJ} \, (dx^I - C^I du) (dx^J - C^J du),
\end{equation}
with the metric functions satisfying the following fall-off conditions at 
large $r$:\footnote{In the previous papers \cite{fakenews, dual0 , dualex}, 
we used slightly stronger fall-off conditions.  Generally, our strategy is 
to assume as weak a set of fall-off conditions as possible 
consistent with our results.}
\begin{align}
 F(u,r,x^I) &= 1 + \frac{F_0(u,x^I)}{r} +  o(r^{-1}), \notag \\[2mm]
 \beta(u,r,x^I) &= \frac{\beta_0(u,x^I)}{r^2} + o(r^{-2}), \notag \\[2mm] 
 C^I(u,r,x^I) &= \frac{C_0^I(u,x^I)}{r^2} + o(r^{-2}), \notag \\[2mm] \label{met:falloff}
 h_{IJ}(u,r,x^I) &= \omega_{IJ} + \frac{C_{IJ}(u,x^I)}{r} + o(r^{-1}),
\end{align}
Furthermore, a gauge freedom allows us to choose
\begin{equation} \label{det:h}
 h =\omega,
\end{equation}
where $h \equiv \textup{det}(h_{IJ})$ and $\omega 
                      \equiv \textup{det}(\omega_{IJ}) =\sin\theta$.
                      
Given the above fall-off conditions for the metric and assuming that all the tensors defined on the 2-sphere above are regular, there exists a new set of dual BMS charges \cite{dual0,dualex}
\begin{align}
 \ndelta \widetilde{\mathcal{Q}}_0 =\frac{1}{16 \pi G} \int_{S}\,  d\Omega \, \Bigg[ \delta \Bigg( - f D_I D_J \widetilde{C}^{IJ}  + \frac{1}{4} Y^K \widetilde{C}^{IJ} D_K C_{IJ} - \frac{1}{2} & \widetilde{Y}^I D_I  C^2 \Bigg) \notag\\
                                                        &\qquad + \frac{1}{2} f \partial_u C_{IJ} \delta \widetilde{C}^{IJ} \Bigg], \label{dual0:Full}
\end{align}
where $D_I$ is the standard covariant derivative associated with the unit round-sphere metric $\omega_{IJ}$, the integration is over the 2-sphere at $r=\infty$ for some fixed $u$, which is 
denoted by $S$, and where 
\begin{equation}
 f= s + \frac{u}{2} D_I Y^I,
\end{equation}
with $s(x^I)$ parameterising a supertranslation and $Y^I(x^J)$ 
corresponding to a conformal Killing vector on the 2-sphere. In
(\ref{dual0:Full}) $C^2$ is equal to $C_{IJ} C^{IJ}$, and the 
twisted quantities, denotated with tildes, are defined as
\begin{equation} \label{Xtwist}
\widetilde{C}^{IJ} = C_K{}^{(I} \epsilon^{J)K}, \quad \widetilde{Y}^I = \epsilon^{IJ} Y_J, \qquad \epsilon_{IJ} =  
\begin{pmatrix}                                                                                0 & 1 \\ -1 & 0                                                              \end{pmatrix} \sin \theta.
\end{equation}
The 0 subscript on $\widetilde{\mathcal{Q}}_0$ denotes that this charge is
the leading-order term in a possible sequence of charges in a $1/r$ expansion near
infinity \cite{dualex}.

These charges complement the Barnich-Troessaert BMS charges \cite{BarTro}\footnote{See equations (3.2) and (3.3) of Ref.\ \cite{BarTro} with the following translations in notation: $mG=-1/2\, F_0$ and $N^I = -3/2 \, C_{1}^I$, where $C_{1}^I$ is an order $1/r^3$ term in the expansion of $C^I$, see equation \eqref{met:falloff}.}
\begin{align}
 \ndelta \mathcal{Q}_0 =\frac{1}{16 \pi G} \int_{S}\, d\Omega \, \Bigg[ \delta \Bigg( - 2 f F_0  + Y^K \Big[ -3 C_{1 K} + \frac{1}{16}   D_K  C^2 \Big]& \Bigg) \notag\\ \label{QBarTro}
                                                        &\quad + \frac{1}{2} f \partial_u C_{IJ} \delta C^{IJ} \Bigg].
\end{align}
One way of understanding the new dual charges is in terms of a complexification of the BMS algebra.  This is made most explicit by writing these expressions in a Newman-Penrose form with the standard BMS charges and the dual charges 
corresponding respectively to the real and imaginary parts of 
appropriate Newman-Penrose scalars \cite{dualex}.

Inspecting equation \eqref{dual0:Full}, it is clear that in order for the dual charges to be non-trivial, at least globally (i.e.\ for $f=1$, $Y^I=0$), we must relax the condition that the metric coefficients be regular tensors on the 2-sphere, in which case they correspond to Taub-NUT charges \cite{dual0, porrati}.  In particular, this means that we must go back through the derivation of both the Barnich-Troessaert and the dual charges and assess whether there are total derivative terms that were previously ignored that will now become relevant due to the relaxing of the regularity conditions.  

However, before we do this in the next section, we make the link with Taub-NUT charges more precise by considering Komar integrals.  For the remainder of the paper, we shall restrict attention to the Abelian part of the BMS algebra given by supertranslations, parameterised by functions $s(x^I)$ on the 2-sphere. 

\subsection{Komar integrals}

Asymptotically flat spacetimes model isolated gravitational systems in general relativity.  The simplest such example is of a point charge of mass/energy $m$, which we obtain by choosing 
\begin{equation}
 F = 1 - \frac{2 m G}{r}, \qquad \beta = 0, \qquad  C^I =0, \qquad h_{IJ} 
= \omega_{IJ}.
\end{equation}
This is, of course, the Schwarzschild solution, written in outgoing Eddington-Finkelstein coordinates.  Given that the solution is stationary, the Komar energy\footnote{We use the expression ``Komar energy'' rather than the more popular ``Komar mass,'' because this makes more sense in the context of this paper.  Of course, for stationary configurations the energy and mass coincide.}
\begin{equation} \label{komar}
 M_{K} = -\frac{1}{8 \pi G} \int_{S} \star d k^\flat
\end{equation}
is well-defined, where $k^\flat$ is the one-form corresponding to 
the timelike Killing vector $k=\partial/\partial u$.  It is a standard exercise to show that 
\begin{equation}
 M_K = m.
\end{equation}

Remaining in the stationary setting for now, it is reasonable to ask whether there exist solutions which have a non-trivial dual Komar energy defined by
\begin{equation} \label{dualKomar}
 \tilde{M}_K = \frac{1}{8 \pi G} \int_{S} d k^\flat
\end{equation}
with the one-form $k^\flat$ again coming from the timelike Killing 
vector $k$, 
as defined above.  The Taub-NUT solution is such an example.  Its metric in 
standard coordinates is \cite{stephani}
\begin{gather}
 ds^2 = - f(r) (dt + 2 \ell \cos \theta d \phi)^2 + f(r)^{-1} dr^2 + (r^2 + \ell^2) (d \theta^2 + \sin ^2 \theta d \phi^2), \notag \\[2mm] \label{met:NUT}
 f(r) = \frac{r^2 - 2mr - \ell^2}{r^2 +\ell^2},
\end{gather}
where $m$ is the Schwarzschild mass parameter as before and $\ell$ is called 
the NUT parameter.  Taking $k^\flat$ to be the one-form coming 
from the Killing vector $k=\partial/\partial t$, it is simple to show that
\begin{equation}
 \tilde{M}_K = \ell/G.
\end{equation}
Thus, the NUT parameter corresponds to a dual gravitational charge.  

An inspection of the metric \eqref{met:NUT} reveals that the Taub-NUT solution is not asymptotically flat, in the sense defined at the start of this section,
where the metric expansion coefficients were implicitly assumed to be
regular on the 2-sphere.\footnote{There is a recent formulation of the asymptotic conditions at spacelike infinity that allows Taub-NUT-like solutions \cite{Henneaux:2018hdj, Henneaux:2019yax}.  However, the extension of these results to null infinity is less clear.}  A simple non-rigorous way of seeing this is that taking the limit as $r \rightarrow \infty$ in metric \eqref{met:NUT} does not lead to the Minkowski metric, precisely because of the existence of the NUT parameter.  A clearer way of seeing this is to write the metric in Bondi coordinates $(u,r,x^I)$, as has been done in Ref.\ \cite{porrati}, and also in appendix B,\footnote{The coordinate transformations performed in 
Ref.\ \cite{porrati} put the metric in Bondi coordinates to the 
necessary order required there.  See appendix \ref{app:TN} 
for further details.} in which case one finds that in particular $C_{IJ}$ is not regular on the 2-sphere.

Given that the Taub-NUT solution is the simplest spacetime with a non-trivial 
NUT charge, this has led to the lore that a NUT parameter is also a measure 
of asymptotic non-flatness,  and that, therefore, \emph{any} spacetime 
with a non-trivial NUT/dual charge is not asymptotically flat (see, however, Ref.\ \cite{Virmani:2011gh}).  Thus, in order to allow for NUT charges in asymptotically flat backgrounds, we need to loosen the definition of asymptotic flatness.  We see this more explicitly by computing the dual Komar energy for a general solution of Bondi form.

Assuming that the general metric \eqref{AF} is stationary with timelike Killing vector $\partial/\partial u$, computing the dual Komar integral gives
\begin{equation} \label{dual:AF}
 \tilde{M}_K = -\frac{1}{8 \pi G} \int_{S}  \partial_I C_{0\, J}\, dx^I 
\wedge dx^J = -\frac{1}{8 \pi G} \int_{S}  d C_{0},
\end{equation}
where $C_{\,0 I} \equiv \omega_{IJ}\, C_0^{J}$.  We observe that if 
$C_{0\,I}$ is not regular on the sphere, the above integral can be non-zero.  
Therefore, we have a non-trivial dual/NUT charge, provided that we are 
prepared to generalise 
the definition of asymptotic flatness so as to allow the $g_{uI}$ components of
the metric to be non-regular on the sphere.

In the general setting, at leading order, the dual supertranslation charge is given by \cite{dual0,dualex}\footnote{See, for example, equation (5.11) of Ref.\ \cite{dual0} and equations (3.5) and (5.5) of Ref.\ \cite{dualex}.}
\begin{equation} \label{dCharge0}
 \tilde{Q}^{\textup{(int)}}_0 = - \frac{1}{16 \pi G} \int_{S} d\Omega\, s\, D_I D_J \tilde{C}^{IJ},
\end{equation}
where $s$ is the supertranslation parameter.  In deriving the above charge integral, it was assumed that the energy-momentum components of the matter content in the null frame adapted to the Bondi coordinates fall-off as \cite{dual0,dualex}
\begin{equation} \label{falloff:matter}
 T_{00} = o(r^{-4}), \qquad T_{0m} = o(r^{-3}).
\end{equation}
We shall continue to assume these fall-off conditions for the energy-momentum 
tensor throughout the paper.  The Einstein equation then implies that
the null-frame components of the Einstein tensor have the fall-offs 

\begin{align}
  G_{00} = o(r^{-4}) &\quad \implies \quad \beta_0 = -\frac{1}{32}\, C^2,  \\
  G_{0m} = o(r^{-3}) &\quad \implies \quad C_0^I = -\half D_J C^{IJ}. \label{eqn:CI0}
\end{align}
Using the definition of $\tilde{C}^{IJ}$ given in equation \eqref{Xtwist} and equation \eqref{eqn:CI0}, the dual charge \eqref{dCharge0} can be written as 
\begin{equation}
  \tilde{Q}^{\textup{(int)}}_0 =  \frac{1}{8 \pi G} \int_{S} s\ d C_{0}. 
\label{tQint}
\end{equation}
Choosing $s=1$, the general expression above reduces to the definition of the dual Komar energy \eqref{dual:AF}, up to an unimportant minus sign.\footnote{In 
hindsight, it might have been more natural in Refs.\ \cite{dual0, dualex} for
us to have defined the dual charges with a relative minus sign.  However, 
for consistency with Refs.\ \cite{dual0,dualex}, we retain the expressions defined therein. \label{ft:minus}}  This is analogous to the case of the standard supertranslation charge $Q_0$ reducing to the Bondi mass upon choosing $s=1$ \cite{BarTro}.

An archetypal example of a solution with non-vanishing dual 
supertranslation charge is the Taub-NUT solution discussed above, 
and more thoroughly in Ref.\ \cite{porrati}.  Reading off from the Bondi
form of the Taub-NUT metric given in appendix B, one has $C_{0}\equiv
C_{0\, I}\, dx^I=-4\ell\, \sin^2\fft{\theta}{2}\,d\phi$, and so the
integral (\ref{tQint}) gives, taking $s=1$,
\be
\widetilde{\mathcal{Q}}_0= -\fft{\ell}{G}\,.\label{tQtn}
\ee
Alternatively, and equivalently, one can obtain the same result by 
reading off the expression for $C_{IJ}$ given in appendix B, and evaluating 
the integral in equation (\ref{dCharge0}).

It fact, in deriving equation \eqref{dCharge0} in Refs.\ \cite{dual0, dualex}, total derivative terms were dropped, because all metric functions were assumed to be regular on the 2-sphere.  However, in the Taub-NUT example considered above, the resulting expression is non-zero precisely because the metric function is non-regular on the 2-sphere.  Therefore, the original expression for the charge is no longer valid in this case and we need to revisit the derivation in Refs.\ \cite{dual0, dualex}.\footnote{The revised dual charge for the Taub-NUT solution is in fact given in equation \eqref{TB:new}.}  This is what we turn to next.

\section{Revisiting BMS charges and the phase space of gravitational modes} \label{sec:phase}

By now it should be clear that the concept of dual BMS charges lends itself better to solutions in which the metric components and more specifically at the leading order in a $1/r$ expansion \cite{fakenews, dualex}, $C_{0 I}$, and 
possibly also $C_{IJ}$, are not regular on the 2-sphere.  Therefore, in the remainder, we shall focus on such spacetimes given by the metric \eqref{AF} with the fall-offs \eqref{met:falloff} and the gauge condition \eqref{det:h}, except that now, we generalise to tensors that are not necessarily regular on the 2-sphere.

The Barnich-Troessaert and dual charges have, however, been derived assuming regularity of the metric components on the 2-sphere.  In practical terms, this means that total derivatives on the 2-sphere can be ignored.  Therefore, in order to proceed, we need to revisit those derivations and define charges appropriate to the solutions in which we are now interested.  Fortunately, this is 
relatively straightforward.  For the usual supertranslation charges, the result can essentially be read off from equation (A.9) of Ref. \cite{BarTro}:
\begin{align}
 Q^{\textup{(int)}}_0 =-\frac{1}{8 \pi G} \int_{S}\, d\Omega \,  \Bigg( s F_0  + \frac{1}{4} D_I (s D_J C^{IJ}) \Bigg),
\end{align}
while the dual supertranslation charges are given by 
\begin{align}
 \widetilde{Q}^{\textup{(int)}}_0 = -\frac{1}{8 \pi G} \int_{S}\, d\Omega \,  \Bigg( \frac{1}{2} s D_I D_J \widetilde{C}^{IJ}  - \frac{1}{4} D_I (s D_J \widetilde{C}^{IJ}) \Bigg),
\end{align}
where we have used equation (4.1) of Ref.\ \cite{dualex} and equation \eqref{eqn:CI0}.  Note that the expressions for the global charges (with $s=1$) change as a result of this rederivation:
\begin{equation}
  Q^{\textup{(int)}}_0(s=1) = -\frac{1}{8 \pi G} \int_{S}\, d\Omega \,  \Bigg( F_0  + \frac{1}{4} D_I D_J C^{IJ} \Bigg),
\end{equation}
instead of $-\fft1{8\pi G}\int_S d\Omega\, F_0$,
and 
\begin{equation}
    \widetilde{Q}^{\textup{(int)}}_0(s=1) = -\frac{1}{32 \pi G} \int_{S}\, 
d\Omega \,  \Bigg( D_I D_J \widetilde{C}^{IJ} \Bigg),
\end{equation}
instead of (\ref{dCharge0}).

The global dual supertranslation charge for the Taub-NUT metric 
has now acquired a factor of $\fft12$ as
a result of our inclusion of the total derivatives that were previously
omitted.  Thus, 
for the Taub-NUT solution we now find
\begin{equation} \label{TB:new}
 \widetilde{Q}^{\textup{(int)}}_0(s=1) = - \fft{\ell}{2G},
\end{equation}
rather than the previous expression (\ref{tQtn}).  

  The fact that the dual charge evaluates to one half of the expected value 
for the Taub-NUT solution may initially seem strange.  This issue 
is reminiscent of the well-known factor of $\fft12$ puzzle with the usual
Komar energy integral (\ref{komar}), which was resolved in 
Ref.~\cite{IW} by noting that the conserved Iyer-Wald charge $Q_{IW}$ 
has two contributions:
\begin{equation} \label{qiw}
 Q_{IW} = \frac{1}{2} M_K - \int_S k \cdot B,
\end{equation}
where $M_K$ is given by (\ref{komar}), $k$ is the timelike Killing 
vector and $B$ is defined by
\begin{equation}
 \delta \int_S \xi \cdot B = \int_S \xi \cdot \Theta.
\end{equation}
Here $\xi$ is a BMS generator 
and $\Theta$ is the symplectic potential form of Iyer 
and Wald \cite{IW}.  The $(-k\cdot B)$ integral on the right hand side of 
equation \eqref{qiw} is non-vanishing and together with $M_K/2$ yields the 
expected value for $Q_{IW}$, effectively giving $M_K$.  (In fact the
integral of $(-k\cdot B)$ exactly cancels the $M_K/2$ term and replaces it
with the standard expression for the ADM mass \cite{IW}.)  
In our case, for comparison, we have that
\begin{equation}
 \tilde{Q}_0^{\textup{(int)}}(s=1) = -\frac{1}{2} \tilde{M}_K.
\end{equation}
The absence of an analogue of the $k\cdot B$ integral term in this dual 
case means that the factor of $1/2$ does not get amended.

We define a total complexified supertranslation charge, 
denoted by a script $Q$ \cite{dualex}:
\begin{equation} \label{Ccharge}
 \mathcal{Q}_0 = Q^{\textup{(int)}}_0 - i \widetilde{Q}^{\textup{(int)}}_0
\end{equation}
and note that in complex coordinates $(z,\bar{z})$ on the sphere given by
\begin{equation}
 z = \cot\fft{\theta}{2}\, e^{i \phi}, 
\end{equation}
so that
\begin{equation}
 ds^2 = 2 \gamma_{z\bar{z}} dz d\bar{z}, \qquad  
\gamma_{z\bar{z}} = \frac{2}{(1+|z|^2)^2},
\end{equation}
\begin{equation} \label{cQ0}
 \mathcal{Q}_0 = -\frac{1}{16 \pi G} \int_{S}\, d\Omega \,  
\Bigg( 2 s F_0 + s\, D_z^2 C^{zz} + D_{\bar{z}} \,s\, 
D_{\bar{z}} C^{\bar{z}\bar{z}} \Bigg).
\end{equation}
In deriving the above equation we have used equation \eqref{Xtwist} 
and the fact that $\epsilon^{z\bar{z}} = i \gamma^{z\bar{z}}$, 
where $\gamma^{z\bar{z}} = 1/\gamma_{z\bar{z}}$ is the $z\bar{z}$ 
component of the inverse metric.  Note that $d\Omega = \sin{\theta} \, d \theta \, d \phi = \gamma_{z \bar{z}} \,  d^2 z.$

We define the charge evaluated at $\scri^+_-$, i.e.\ on $\scri^+$
at $u= -\infty$, as
\begin{equation}
 \mathcal{Q}^+ = -\frac{1}{16 \pi G} \int_{\scri^+_-}\, d\Omega \,  
\Bigg( 2 s F_0 + s\, D_z^2 C^{zz} + D_{\bar{z}} \,s\, D_{\bar{z}} 
C^{\bar{z}\bar{z}} \Bigg),
\end{equation}
and we assume that 
\begin{equation} \label{Q++}
 \mathcal{Q}_0|_{\scri^+_+} = \mathcal{Q}_0\Big|_{u=+\infty} = 0.
\end{equation}
Note from equations \eqref{dual0:Full} and \eqref{QBarTro} that the flux for the total charge is controlled by $\partial_u C_{IJ}.$\footnote{The non-integrable pieces for both charges remain the same when dropping the requirement that the metric components be regular on the 2-sphere.}  Thus, $\mathcal{Q}^+$ may be written as
\begin{equation}
 \mathcal{Q}^+ = \frac{1}{16 \pi G} \int_{\scri^+}\, du d\Omega \,  
\Bigg( 2 s\, \partial_u F_0 + s\, D_z^2 N^{zz} + D_{\bar{z}} \,s\, 
D_{\bar{z}} N^{\bar{z}\bar{z}} \Bigg),
\end{equation}
where $N_{zz} = \partial_u C_{zz}.$
Assuming for simplicity that 
\begin{equation}
 T_{11} = o(r^{-2}),
\end{equation}
the Einstein equation gives that
\begin{equation} \label{uF0}
 \partial_u F_0 = - \frac{1}{2} D_I D_J N^{IJ} + \frac{1}{4} N^{IJ} N_{IJ} = - \frac{1}{2} (D_z^2 N^{zz} + D_{\bar{z}}^2 N^{\bar{z}\bar{z}}) + \frac{1}{2} N^{zz} N_{zz},
\end{equation}
which implies that
\begin{equation} \label{Q+}
 \mathcal{Q}^+ = \frac{1}{16 \pi G} \int_{\scri^+}\, du\, d^2 z\, 
\gamma^{z\bar{z}} \,  \Bigg( s N_{zz} N_{\bar{z}\bar{z}} - s D_{\bar{z}}^2 N_{zz} + D_{\bar{z}} s\, D_{\bar{z}} N_{zz} \Bigg).
\end{equation}
We have put the charge into a form in which we can investigate how it acts on gravitational modes given by $C_{zz}$ and $C_{\bar{z}\bar{z}}$.  The Dirac bracket for the radiative modes at null infinity is \cite{Ashtekar:1981bq,Ashtekar:1987tt}
\begin{equation}
 \{N_{zz}(u,z,\bar{z}), N_{\bar{w}\bar{w}}(u',w,\bar{w})\} = - 16 \pi G\, \partial_u \delta(u-u')\, \delta^2(z-w)\, \gamma_{z\bar{z}}, 
\end{equation}
which implies that 
\begin{equation}
 \{N_{zz}(u,z,\bar{z}), C_{\bar{w}\bar{w}}(u',w,\bar{w})\} = 16 \pi G\, \delta(u-u')\, \delta^2(z-w)\, \gamma_{z\bar{z}}.
\end{equation}
Now, computing the bracket of $\mathcal{Q}^+$ with $C_{zz}$ and with 
$C_{\bar{z}\bar{z}}$ gives
\begin{align}
 &\{\mathcal{Q}^+, C_{zz}(u,z,\bar{z})\} = s \, \partial_u C_{zz}, \\
 &\{\mathcal{Q}^+, C_{\bar{z}\bar{z}}(u,z,\bar{z})\} = s \, \partial_u C_{\bar{z}\bar{z}} - 2 \, D_{\bar{z}}^2 s,
\end{align}
where we have freely integrated by parts in terms involving ${\delta^2(z-w)}$.  Thus, $\mathcal{Q}^+$ generates time translation on $C_{zz}$, while it generates a supertranslation on $C_{\bar{z}\bar{z}}$.  This is in complete analogy with electromagnetism, where $A^{(0)}_z$ transforms as a gauge parameter under a complexified U(1) charge, while $A^{(0)}_{\bar{z}}$ is invariant under its action \cite{Strom:mag}.\footnote{Note that the action of $\mathcal{Q}^+$ on $C_{zz}$ and $C_{\bar{z}\bar{z}}$ would have been reversed had we defined $\widetilde{\mathcal{Q}}^{\textup{(int)}}_0$ with a relative minus sign (see footnote \ref{ft:minus}); or equivalently, had we considered the complex conjugate of $\mathcal{Q}.$ }  Thus, we find that the phase space problem that was identified in Ref.\ \cite{Strom:soft2}, namely, that the action of the supertranslation charge on $C_{zz}$ was incorrect, is resolved here by considering the \emph{full} complexified supertranslation charge, i.e. the usual supertranslation charge, as well as the dual one.  In Ref.\ \cite{Strom:soft2}, this problem was resolved by imposing a restriction on phase space that effectively proscribed dual charges.  From the perspective taken here, this makes sense in that if the dual charge is to be assigned a zero value, then this must be reflected and enforced on the phase space.  In Ref.\ \cite{porrati}, a resolution of this problem is achieved by imposing a dyonic boundary condition on the phase space.  However, there is \textit{a priori} no reason to expect that boundary conditions should be needed in general.\footnote{We expect that the requirement that the total charge vanishes at $\scri^+_+$, see equation \eqref{Q++}, is a technical one and may be removed by a better understanding of the phase space.   In any case, from the perspective of
an initial value problem, 
we are always free to choose such a boundary condition.}  Moreover, from the perspective of the initial value problem, prescribing boundary conditions at both ends of future null infinity is not entirely satisfactory.

\section{Soft NUT/graviton theorem} \label{sec:soft}

Electromagnetism and general relativity are gauge theories, in the sense that there are redundant U(1) and diffeomorphism transformations, respectively, which leave the physics unchanged.  In the context of asymptotically flat spacetimes, which form a class of solutions in general relativity defined by specific boundary conditions, there exists an asymptotic symmetry group, the BMS group, (or more precisely an algebra) that acts in such a way as to preserve the boundary conditions.  Specifically, the action of a BMS generator on an asymptotically flat metric gives back another asymptotically flat metric.  However, given that the BMS group acts at null infinity, it induces large gauge transformations, which cannot be viewed as redundancies.  Thus, BMS transformations are physically relevant transformations.  

From a classical scattering point of view, conservation of charge ${\bf{Q}}$ translates to the fact that
\begin{equation}
 {\bf{Q}}^+ = {\bf{Q}}^-,
\end{equation}
where ${\bf{Q}}^+ = {\bf{Q}}|_{\scri^+_-}$ and  
${\bf{Q}}^- = {\bf{Q}}|_{\scri^-_+}$ are the limiting values of the relevant charge ${\bf{Q}}(u,x^I)$.  In the quantum theory this becomes an operator identity
\begin{equation}
 {\bf{Q}}^+ \, S - S\, {\bf{Q}}^- = 0.
\end{equation}
Such an identity has so far been conjectural \cite{Strom:YM, Strom:soft1}, but should ultimately come from the theory.\footnote{See Refs.~\cite{Henneaux:2018gfi,Henneaux:2018hdj,Henneaux:2019yax} for investigations in this direction.}  In any case, the remarkable observation \cite{Strom:soft2, Strom:QED} (see Ref.~\cite{Strom:lec} for a review) is that such a conservation equation is equivalent to the Weinberg soft photon and graviton theorems in the context of electromagnetism and general relativity.  Moreover, considering magnetic monopole charges leads to a Weinberg type soft magnetic monopole theorem \cite{Strom:mag}.  

In this section, we derive a soft NUT/graviton theorem using the complexified supertranslation charges found in section \ref{sec:phase}, see equation \eqref{Ccharge}, for asymptotically flat spacetimes generalised to include non-regular tensors on the 2-sphere.  We introduce Bondi coordinates $(v,r,x^I)$ adapted to ingoing null geodesics so that the metric takes the form
\begin{equation} \label{vAF}
 d s^2 = - \hat{F} e^{2 \hat{\beta}} dv^2 + 2 e^{2 \hat{\beta}} dv dr + 
r^2 \hat{h}_{IJ} \, (dx^I - \hat{C}^I dv) (dx^J - \hat{C}^J dv)
\end{equation}
with the metric components satisfying the same fall-off conditions as with outgoing Bondi coordinates given by \eqref{met:falloff}, where we denote all respective objects on the right hand side with a hat, and the determinant condition \eqref{det:h}.

We begin with the conservation of charge equation
\begin{equation}
\mathcal{Q}^+ = \mathcal{Q}^-,
 \label{nutsoft}
\end{equation}
where $\mathcal{Q}^+$ is given by equation \eqref{Q+} and $\mathcal{Q}^-$ 
corresponds to the charge at $\scri^-_+$ associated with a supertranslation 
parameter $\hat{s}$,
\begin{align}
 \mathcal{Q}^- &= -\frac{1}{16 \pi G} \int_{\scri^-_+}\, d\Omega \,  \Bigg( 2 \hat{s} \hat{F}_0 + \hat{s}\, D_z^2 \hat{C}^{zz} + D_{\bar{z}} \hat{s}\, D_{\bar{z}} \hat{C}^{\bar{z}\bar{z}} \Bigg) \\
 &= \frac{1}{16 \pi G} \int_{\hat{\scri^-}}\, dv\, d^2 z\, \gamma^{z\bar{z}} \,  \Bigg( \hat{s} \hat{N}_{zz} \hat{N}_{\bar{z}\bar{z}} + \hat{s} D_{\bar{z}}^2 \hat{N}_{zz} - D_{\bar{z}} \hat{s}\, D_{\bar{z}} \hat{N}_{zz} \Bigg),
\end{align}
where in the 
second line we have used the fact that the charge vanishes at $\scri_-^-$ and that
\begin{equation}
  \partial_v \hat{F}_0 = - \frac{1}{2} (D_z^2 \hat{N}^{zz} + D_{\bar{z}}^2 \hat{N}^{\bar{z}\bar{z}}) + \frac{1}{2} \hat{N}^{zz} \hat{N}_{zz}.
\end{equation}

Now, we would like to split the charges into so-called hard and soft parts; namely, a part related to flux at future/past null infinity that reduces/increases the charge and a part that is related to the large diffeomorphisms.  We determine the hard part of the charge by considering the flux formula for the global charge.  Taking the $u$-derivative of the global charge defined on future null infinity given in equation \eqref{cQ0} with $s=1$
\begin{align}
 \partial_u \mathcal{Q}_0 (s=1) &= -\frac{1}{16 \pi G} \int_{S}\, d\Omega \,  \Bigg( 2 \partial_u F_0 + \, D_z^2 N^{zz}  \Bigg) \\
                                &= -\frac{1}{16 \pi G} \int_{S}\, d\Omega \,  \Bigg( N^{zz} N_{zz} - \, D_{\bar{z}}^2 N^{\bar{z}\bar{z}}  \Bigg),
\end{align}
where in the second line we have used equation \eqref{uF0}, we define the hard charge to be
\begin{equation}
 \mathcal{Q}^{+}_H = \frac{1}{16 \pi G} \int_{\scri^+}\, 
du \, d^2 z\, \gamma^{z \bar{z}} \, s \Bigg( N_{zz} N_{\bar{z}\bar{z}} - \, D_{\bar{z}}^2 N_{zz}  \Bigg)
\end{equation}
so that, from equation \eqref{Q+}
\begin{equation}
 \mathcal{Q}^+ = \mathcal{Q}^{+}_H + \mathcal{Q}^{+}_S
\end{equation}
with
\begin{equation}
 \mathcal{Q}^{+}_S = \frac{1}{16 \pi G} \int_{\scri^+}\, du\, d^2 z\, \gamma^{z\bar{z}} \,   D_{\bar{z}} s\, D_{\bar{z}} N_{zz}.
\end{equation}

Similarly, 
\begin{equation}
  \mathcal{Q}^- = \mathcal{Q}^{-}_H + \mathcal{Q}^{-}_S,
\end{equation}
where
\begin{equation}
 \mathcal{Q}^{-}_H = \frac{1}{16 \pi G} \int_{\scri^-}\, dv\, d^2 z\, \gamma^{z\bar{z}} \, \hat{s} \Bigg(  \hat{N}_{zz} \hat{N}_{\bar{z}\bar{z}} +  D_{\bar{z}}^2 \hat{N}_{zz} \Bigg)
\end{equation}
and 
\begin{equation}
 \mathcal{Q}^{-}_S = - \frac{1}{16 \pi G} \int_{\scri^-}\, dv\, d^2 z\, \gamma^{z\bar{z}} \, D_{\bar{z}} \hat{s}\, D_{\bar{z}} \hat{N}_{zz}.
\end{equation}
Setting $s=\hat{s}$, which amounts to breaking the BMS$^+ \times$BMS$^-$ 
symmetry to its diagonal subgroup \cite{Strom:soft1}, we obtain a 
complexified supertranslation Ward identity,
\begin{equation} \label{Ward}
 \mathcal{Q}^+ S - S \mathcal{Q}^- = 0.
\end{equation}

We consider as scattering states particles of energy $E$ and NUT charge 
$\tilde{E}.$ For example, such  configurations could be Lorentzian
signature multi-NUT solutions 
\cite{Bossard:2008sw}.
Identifying the action of $\mathcal{Q}^-$ on an $n$-particle state $| z^{in}_1, \ldots z^{in}_n \rangle$ to be
\begin{equation}
 \mathcal{Q}^- | z^{in}_1, \ldots, z^{in}_n \rangle = \sum_{k=1}^n (E^{in}_k - i \widetilde{E}^{in}_k)\, s(z^{in}_{k}) \ | z^{in}_1, \ldots, z^{in}_n \rangle
\end{equation}
and similarly,
\begin{equation}
 \langle z^{out}_1, \ldots, z^{out}_m |\mathcal{Q}^+  =  \sum_{k=1}^m (E^{out}_k - i \widetilde{E}^{out}_k)\, s(z^{out}_{k}) \ \langle z^{out}_1, \ldots, z^{out}_m |,
\end{equation}
the Ward identity \eqref{Ward}, sandwiched between the ingoing and outgoing states simplifies to
\begin{samepage}
\begin{align}
  \langle z^{out}_1,& \ldots, z^{out}_m | \mathcal{Q}^{+}_S S - S \mathcal{Q}^{-}_S | z^{in}_1, \ldots, z^{in}_n \rangle  \\[2mm] \notag
 &= \Bigg[ \sum_{k=1}^n (E^{in}_k - i \widetilde{E}^{in}_k) s(z^{in}_{k}) - \sum_{k=1}^m (E^{out}_k - i \widetilde{E}^{out}_k) s(z^{out}_{k})\Bigg] \langle z^{out}_1, \ldots, z^{out}_m | S | z^{in}_1, \ldots, z^{in}_n \rangle.
\end{align}
\end{samepage}
Choosing $s(w) = \frac{1}{z-w}$ \cite{Strom:soft2}, we obtain a Ward identity that corresponds to a new soft NUT/graviton theorem.

\section*{Acknowledgements}

We would like to thank Mihalis Dafermos, Malcolm Perry, Andrew Strominger, Cedric Troessaert and Alexander Zhiboedov for useful discussions.  We would like to thank the Mitchell Family Foundation for hospitality at the 2019 Cook's Branch workshop. Moreover, M.G.\ and C.N.P.\ would like to thank the Max-Planck-Institut f\"ur Gravitationsphysik (Albert-Einstein-Institut), Potsdam and H.G.\ would like to thank Queen Mary University of London for hospitality during the course of this work.  M.G.\ is supported by a Royal Society University Research Fellowship. C.N.P.\ is partially supported by DOE grant DE-FG02-13ER42020.

\appendix

\section{Kerr Metric in BMS Coordinate Gauge}

  In this appendix we shall give a construction of a Bondi coordinate
system for the Kerr metric, up to the first few orders in a $1/r$ 
expansion.  Following this, in appendix B, we give an analogous
construction of the Taub-NUT metric in Bondi coordinates. One of the 
reasons for doing this is to highlight a key difference between
a metric such as Kerr, for which all the global dual charges vanish, and
a metric such as Taub-NUT, which has non-vanishing global dual charges.  
This difference is reflected in the fact that whereas the various 
scalar, vector and
tensor fields in the expansion of the Bondi form of the Kerr metric are
all non-singular on the 2-sphere,\footnote{A construction of a Bondi
form for the Kerr metric was presented in \cite{BarTro}, and many of the
fields in the expansion of the metric did have singularities at the poles of
the sphere.  This, however, was an artefact of the coordinates that were
used in \cite{BarTro}: using asymptotically spheroidal 
rather than asymptotically spherical coordinates.  A construction of
the Kerr metric in Bondi coordinates had been described previously
in \cite{fletlun}, but not in a gauge that is convenient for our
purposes.}
 many of the analogous fields in the 
expansion of the Taub-NUT metric are singular on the sphere.

  We take as the starting point the Kerr metric in Boyer-Lindquist 
coordinates:
\bea
ds^2 &=& -\Big(1 -\fft{2 m \bar r}{\bar\rho^2}\Big)\, d\bar t^2 -
  \fft{4 a m \bar r\sin^2\bar\theta}{\bar\rho^2} \, d\bar t d\bar\phi +
  \fft{\bar\rho^2}{\bar\Delta}\, d\bar r^2 + 
  \bar\rho^2\, d\bar\theta^2\nn\\
&& + 
  \Big(r^2+a^2+ \fft{2 a^2 m \bar r \sin^2\bar\theta}{\bar\rho^2}\Big)\,
\sin^2\bar\theta\, d\bar\phi^2\,,\label{KerrBL}
\eea
where
\be
\bar\rho^2 = \bar r^2 + a^2 \cos^2\bar\theta\,,\qquad
 \bar\Delta = \bar r^2+ a^2 - 2m\bar r\,.
\ee
Setting $m=0$, this describes 
Minkowski spacetime in a spheroidal coordinate system, with the metric
\be
ds_{\rm Mink}^2 = 
  -dt^2 + \fft{(\bar r^2 + a^2\, \cos^2\bar\theta^2)\, d\bar r^2}{
(\bar r^2 + a^2)} + (\bar r^2 + a^2\, \cos^2\bar\theta^2) d\bar\theta^2+
(\bar r^2+a^2)\, \sin^2\bar\theta\, d\bar\phi^2\,.
\label{Mink0}
\ee
In order to construct a smooth Bondi coordinate system for
the Kerr metric, we should first transform
to genuine spherical polar coordinates, so that the
spatial metric at large radius will have the desired form, described as 
a foliation of round 2-spheres. This is effected by replacing the 
coordinates $(\bar r, \bar\theta)$ in (\ref{KerrBL}) by 
$(\tilde r,\tilde\theta)$, where
\be
\tilde r^2\,\sin^2\tilde\theta =(\bar r^2 + a^2)\, \sin^2\bar\theta\,,\qquad
\tilde r^2\,\cos^2\tilde\theta = \bar r^2\,\cos^2\bar\theta\,.\label{spheroidal}
\ee
If one sets $m=0$, then under these redefinitions the metric (\ref{Mink0}) 
takes on the standard Minkowski form
\be
ds^2_{\rm Mink}=-d\bar t^2 + d\tilde r^2 + \tilde r^2\, (d\tilde\theta^2 + 
  \sin^2\tilde\theta\, d\bar\phi^2)\,.
\ee

   Since, for our purposes, we eventually want to construct a large-distance 
expansion for the Kerr metric in Bondi coordinates, it will 
suffice at this stage to re-express (\ref{spheroidal}) in the form of a
perturbative large-$\tilde r$ expansion:
\bea
\bar r&=& \tilde r -\fft{a^2\, \sin^2\tilde\theta}{2\tilde r} +
    \fft{a^4\,(3{+}5\cos2\tilde\theta)\sin^2\tilde\theta}{16 \tilde r^3}
 -\fft{a^6\,(15+28\cos2\tilde\theta + 
            21 \cos4\tilde\theta)\sin^2\tilde\theta}{
128 \tilde r^5} +{O}(\tilde r^{-7})\,,\nn\\
\bar\theta &=& \tilde\theta -\fft{a^2\, \sin2\tilde\theta}{4\tilde r^2} +
 \fft{3 a^4\,\sin4\tilde\theta}{32\tilde r^4} -
  \fft{5 a^6\, \sin6\tilde\theta}{96\tilde r^6} +{O}(\tilde r^{-8})\,.
\eea
After making these transformations, the first few terms in the 
Kerr metric (\ref{KerrBL}) take the form
\bea
ds^2 &=&-d\bar t^2 + d\tilde r^2 + \tilde r^2\, (d\tilde\theta^2 + 
  \sin^2\tilde\theta\, d\bar\phi^2) 
 + \fft{2m}{\tilde r}\, \Big[d\tilde r^2 + 
    (d\bar t -a \sin^2\tilde\theta\, d\bar\phi)^2\Big]\nn\\
&& +
 \fft{4m}{\tilde r^2}\, (m d\tilde r -a^2 \sin\tilde\theta\cos\tilde\theta\,
  d\tilde\theta)^2 + O(\tilde r^{-3})\,.
\eea
Here we display only a few terms; we actually worked to a sufficiently 
high order in the expansion for our later purposes. 

  We then look for a further transformation to coordinates $(u,r,\theta,\phi)$ 
in which the Kerr metric takes the Bondi form. We do this by considering 
large-$r$ perturbative expansions of the form
\bea
\bar t&=& u + c\, \log r + h_0(\theta)\, r + h_1(\theta)\, + 
\fft{h_2(\theta)}{r} + \fft{h_3(\theta)}{r^2} + \fft{h_4(\theta)}{r^3} 
 + \cdots\,,\nn\\
\bar\phi &=& \phi + \alpha_0(\theta) + \fft{\alpha_1(\theta)}{r} + 
  \fft{\alpha_2(\theta)}{r^2} + \fft{\alpha_3(\theta)}{r^3} + \cdots\,,\nn\\
\tilde r&=& r + g_0(\theta) + \fft{g_1(\theta)}{r} + \fft{g_2(\theta)}{r^2} +
  \fft{g_3(\theta)}{r^3} + \cdots\,,\nn\\
\tilde\theta &=& \theta + \fft{\gamma_0(\theta)}{r} +
\fft{\gamma_1(\theta)}{r^2} + \fft{\gamma_2(\theta)}{r^3} +\cdots\,,
\label{Bondianz}
\eea
where $c$ is an as-yet undetermined constant and $h_i$, $\alpha_i$, $g_i$ and
$\beta_i$ are as-yet undetermined functions of $\theta$.
We then determine 
these quantities by imposing the requirements, order by order in powers of
$1/r$, that
\be
g_{rr}=0\,,\qquad g_{r\theta}=0\,,\qquad g_{r\phi}=0\,,\qquad
  \det(g_{IJ}) = r^4 \, \det(\omega_{IJ})\,,\label{Bondicon}
\ee
where $\omega_{IJ}$ is the standard metric on the unit 2-sphere (and
so $\det(\omega_{IJ})=\sin^2\theta$).  

   There is some freedom in the choice of the functions in the coordinate
transformations, which reflects the fact that the BMS group maps one 
choice into another.  Up to the first few orders, a choice of BMS gauge that 
achieves the conditions (\ref{Bondicon}) is to take\footnote{The delayed
onset of the $r$-dependent terms in the expansion for
$\tilde\theta$ in (\ref{Bondianz}) and 
(\ref{Bondicoords}) is one manifestation of the 
freedom in choosing a BMS gauge.}
\crampest
\bea
\bar t &=& u+r + 2m \log r -\fft{4m^2}{r} + \fft{m(a^2 + 3 a^2\, \cos2\theta
- 16 m^2)}{4 r^2} 
+\fft{m^2\, (7a^2 + 9 a^2\, \cos2\theta - 32 m^2)}{6 r^3}
 +\cdots\,,\nn\\
\bar\phi &=& \phi - \fft{m a}{r^2} - \fft{4 m^2\, a}{3 r^3} +
  \fft{m a\, (3a^2+ 5 a^2\, \cos 2\theta - 16 m^2)}{8 r^4} +\cdots\,,\nn\\
\tilde r &=& r -\fft{m a^2\, \sin^2\theta}{2 r^2} +
    \fft{m a^4\, (3+5\cos 2\theta)\sin^2\theta}{8 r^4} +
 \cdots\,,\nn\\
\tilde\theta &=& \theta + \fft{m a^4\, \cos\theta\, \sin^3\theta}{4 r^5}
-\fft{m a^6\, (5\cos\theta + 3 \cos3\theta)\, \sin^3\theta}{16 r^7}+\cdots
\,.\label{Bondicoords}
\eea
\uncramp
Up to the order we have calculated, 
the components of the Kerr metric in these 
Bondi coordinates are given by
\crampest
{\setlength{\jot}{10pt}
\bea
g_{uu} &=& -1 + \fft{2m}{r} - \fft{m a^2\, (1{+}3\cos2\theta)}{2 r^3} +
  \fft{m^2\, a^2\, \sin^2\theta}{r^4} +
\fft{m a^4\, (9 {+} 20 \cos 2\theta {+} 35 \cos 4\theta)}{32 r^5} +
O(r^{-6})\,,\nn\\
g_{ur} &=& -1 + O(r^{-6})\,,\nn\\
g_{u\theta} &=& 
  \fft{3 m a^2\, \sin\theta\, \cos\theta}{r^2} -
   \fft{5m a^4\, (2 \sin2\theta + 7 \sin 4\theta)}{32 r^4} +
   \fft{2 m^2\, a^4\, \cos\theta\, \sin^3\theta}{r^5}+O(r^{-6})\,,\nn\\
g_{u\phi} &=& -\fft{2m a \sin^2\theta}{r} +
  \fft{m a^3\,(3 + 5 \cos2\theta)\sin^2\theta}{2 r^3} -
   \fft{ m^2\, a^3\, \sin^4\theta}{r^4} \nn\\
&& - \fft{3 m a^5\, (15 + 28 \cos2\theta +
21 \cos4\theta)\sin^2\theta}{32 r^5} + O(r^{-6})\,,\nn\\
g_{rr} &=&O(r^{-7})\,,\qquad g_{r\theta}=O(r^{-7}) 
 \,,\qquad g_{r\phi}=O(r^{-7})\,,\nn\\
g_{\theta\theta}&=& r^2 - \fft{m a^2\, \sin^2\theta}{r} + 
 \fft{3m a^4\, (5+7 \cos2\theta)\sin^2\theta}{8 r^3} + O(r^{-5})\,,\nn\\
g_{\theta\phi} &=& -\fft{5 m a^3\, \cos\theta\sin^3\theta}{2 r^2} 
  + \fft{7m\, a^5\, (5\cos\theta + 3 \cos3\theta)\, \sin^3\theta}{8r^4} 
+O(r^{-5})\,,\nn\\
g_{\phi\phi} &=& r^2\, \sin^2\theta+ \fft{m a^2\, \sin^4\theta}{r} 
 - \fft{3m a^4\, (5+7\cos2\theta)\sin^4\theta}{8 r^3}+ 
O(r^{-5})\,.
\eea
}
\uncramp

\vspace*{-10mm}

\noindent Comparing with the expansions of the Bondi metric coefficients, as
defined in \cite{fakenews}, we see in particular
that
\bea
C_{IJ} & =& 0\,,\nn\\
D_{IJ}:&& 
  D_{\theta\theta} =-m a^2\, \sin^2\theta\,,\qquad D_{\theta\phi}=0\,,\qquad
  D_{\phi\phi} = m a^2\, \sin^4\theta\,,\nn\\
E_{IJ}:&& E_{\theta\theta}= 0\,,\qquad 
E_{\theta\phi}=- 5 m a^3\, \cos\theta\, \sin^3\theta\,,\qquad 
E_{\phi\phi}=0\,,\nn\\
C_{0\, I}& =&0 \,,\nn\\
C_{1\, I}:&& C_{1\, \theta}=0\,,\qquad C_{1\, \phi}= 2 m a\, \sin^2\theta\,.
\eea
It should be noted that neither these, nor any of the other components of
the metric, are singular at the poles, or anywhere else, on the sphere.

\section{Taub-NUT Metric in Bondi Coordinates}

\label{app:TN}

  Here, we construct the first few orders in the expansion of the 
Taub-NUT metric in Bondi coordinates.  This is similar in spirit to our 
expansion for the Kerr metric in Bondi coordinates, except that here the fact
that Taub-NUT is not globally asymptotically flat inevitably means that
there will be singularities in some of the metric coefficients at one
or more locations on the sphere. 
We begin by sending $t\rightarrow \bar t- 2\ell \phi$ 
in the metric
(\ref{met:NUT}), so that the wire singularity occurs 
only at the south pole of the
sphere:
\be
ds^2 = -f(\bar r)\,(d\bar t-4\ell\,\sin^2\ft{\bar\theta}{2})\,d\bar\phi)^2 
+f(\bar r)^{-1}\,d\bar r^2+
   (\bar r^2\ell^2)\,(d\bar\theta^2+\sin^2\, \bar\theta\,d\bar\phi^2)\,.
\label{TNmetsouth}
\ee
We have placed bars on the coordinates, because we now make an expansion
of the form (\ref{Bondianz}), imposing the Bondi metric conditions 
(\ref{Bondicon}) order by order in the expansion in $1/r$.  Proceeding to
the first few order, we find 
{\setlength{\jot}{10pt}
\crampest
\bea
\bar t&=& u+r + 2m \log r + \fft{\ell^2\,(4+3\cos\theta)\sec^4\fft{\theta}{2}
- 8m^2-11\ell^2}{2r}
+ \fft{m[ \ell^2\, \sec^4\fft{\theta}{2} 
 -4(\ell^2 + m^2)]}{r^2} +\cdots\,,\nn\\
\bar\phi&=& \phi +\fft{\ell \sec^2\fft{\theta}{2}}{r} +
\fft{\ell^3\, (3\cos2\theta + 12\cos\theta -31)\, \sec^6\fft{\theta}{2}}{
     48 r^3}  +\fft{m\,\ell^3 \,\sin^2\fft{\theta}{2}\,\sec^4\fft{\theta}{2}}{
                        r^4} +\cdots\,,\nn\\
\bar r &=& r + \fft{\ell^2\,(3\cos\theta+5)\,\sin^2\fft{\theta}{2}\,
\sec^4\fft{\theta}{2}}{4r} 
- \fft{2m\ell^2\,\tan^2\fft{\theta}{2}}{r^2}\\
&&
-\fft{\ell^4\,(15\cos3\theta + 70\cos2\theta + 225\cos\theta + 202)\,
\sin^2\fft{\theta}{2}\, \sec^8\fft{\theta}{2}}{256 r^3} +
\fft{2m\ell^4 \,\sin^2\fft{\theta}{2}\,\sec^6\fft{\theta}{2}}{r^4}+\cdots\,,
\nn\\
\bar\theta &=& \theta  -\fft{\ell^2\,\sin\fft{\theta}{2}\,
                 \sec^3\fft{\theta}{2}}{r^2} -
\fft{\ell^4\,(\cos2\theta + 2\cos\theta-9)\, \sin\fft{\theta}{2}\, 
  \sec^7\fft{\theta}{2}}{8r^4} -
\fft{2m\ell^4\,\sin^3\fft{\theta}{2} \,\sec^5\fft{\theta}{2}}{r^5}+\cdots\,.\nn
\eea
\uncramp
}

\vspace*{-10mm}

\noindent We have actually worked to a higher order than the terms presented here,
sufficient for our later purposes.  
Using these expansions, we then obtain the Taub-NUT metric in Bondi form,
finding
{\setlength{\jot}{10pt}
\bea
g_{uu} &=& -1 +\fft{2m}{r} + \fft{2\ell^2}{r^2} +
\fft{m\ell^2 \, (\cos2\theta - 4\cos\theta -13)\,\sec^4\fft{\theta}{2}}{8r^3}
\nn\\
&&
+ \fft{[(\ell^2-m^2)\cos2\theta+ m^2 -5 \ell^2\,]\sec^4\fft{\theta}{2}}{2 r^4}
  +{\cal O}(r^{-5})\,,\nn\\
g_{ur} &=& -1 +\fft{\ell^2\,\tan^4\fft{\theta}{2}}{2 r^2}+
\fft{3\ell^4\,(7\cos2\theta {+} 36\cos\theta  {+} 21)\, \sin^4\fft{\theta}{2}\,
\sec^8\fft{\theta}{2}} {64r^4} -\fft{8m\ell^4\tan^4\fft{\theta}{2}}{r^5}+
{\cal O}(r^{-6})\,,\nn\\
g_{u\theta}&=& -\fft{\ell^2\,(\cos2\theta+3\cos\theta+4)\sin\fft{\theta}{2}\,
\sec^5\fft{\theta}{2}}{2r} +\fft{2m\ell^2\, (2\cos\theta+1)\,
  \sin\fft{\theta}{2}\,\sec^3\fft{\theta}{2}}{r^2} \nn\\
&&+
\fft{\ell^4\,(3\cos4\theta+27\cos3\theta+114\cos2\theta+213\cos\theta+155)\,
\sin\fft{\theta}{2}\,\sec^9\fft{\theta}{2}}{64 r^3} \nn\\
&&+
\fft{m \ell^4\,(3\cos3\theta -4\cos2\theta -51\cos\theta-28)\,
\sin\fft{\theta}{2}\, \sec^7\fft{\theta}{2}}{8 r^4}+{\cal O}(r^{-5})\,,\nn\\
g_{u\phi} &=& 4\ell\,\sin^2\fft{\theta}{2} - 
\fft{8m\ell \sin^2\fft{\theta}{2}}{r}-\fft{4\ell^3\,(\cos\theta+2)\,
 \tan^2\fft{\theta}{2}}{r^2}\nn\\
&& -
\fft{m \ell^3\,(\cos2\theta-12\cos\theta-21)\sin^2\fft{\theta}{2}\,
               \sec^4\fft{\theta}{2}}{2 r^3} \nn\\
&&\!\!\!\!\! \!\!\!\!\!\!  -
\fft{\ell^3\,[(\ell^2 {-} m^2)\cos3\theta +(\ell^2 {-} 2 m^2)\cos2\theta -
   (17\ell^2 {-} m^2)\cos\theta {+} 2m^2 {-} 25\ell^2]\sin^2\fft{\theta}{2}\,
   }{2r^4\, \cos^6\fft{\theta}{2}} 
    + {\cal O}(r^{-5})\,,\nn\\
g_{rr}&=& {\cal O}(r^{-6})\,,\qquad g_{r\theta}={\cal O}(r^{-5})\,,\qquad
g_{r\phi}= {\cal O}(r^{-6})\,,\nn\\
g_{\theta\theta}&=& r^2 + 2\ell^2\, \tan^4\fft{\theta}{2} -
 \fft{4m\ell^2\,\tan^2\fft{\theta}{2}}{r} -
 \fft{\ell^4\,(9\cos\theta+1)\,\sin^2\fft{\theta}{2}\,\sec^4\fft{\theta}{2}}{
  r^2}\nn\\
&& +\fft{m\ell^4\, (15\cos2\theta+28\cos\theta +29)\, 
\sin^2\fft{\theta}{2}\,\sec^6\fft{\theta}{2}}{4r^3} + {\cal O}(r^{-4})\,,\nn\\
g_{\theta\phi}&=& 4\ell r\,\sin^3\fft{\theta}{2}\,\sec\fft{\theta}{2} +
  \fft{\ell^3\, (5\cos2\theta + 12\cos\theta+15)\,\sin^3\fft{\theta}{2}\,
    \sec^5\fft{\theta}{2}}{4r} -\fft{20m \ell^3\,\sin^3\fft{\theta}{2}\,
  \sec\fft{\theta}{2}}{r^2} \nn\\
&&-
\fft{\ell^5\,(17\cos4\theta +376 \cos3\theta+1884\cos2\theta +
4168\cos\theta {+} 2771)\,\sin^3\fft{\theta}{2}\,\sec^9\fft{\theta}{2}}{
   256 r^3} + {\cal O}(r^{-4})\,,\nn\\
g_{\phi\phi}&=& r^2\,\sin^2\theta +8\ell^2\, \sin^6\fft{\theta}{2}\,
\sec^2\fft{\theta}{2} + \fft{16 m\ell^2\, \sin^4\fft{\theta}{2}}{r}+
\fft{4\ell^4\,(\cos\theta+9)\,\sin^4\fft{\theta}{2}\,
\sec^2\fft{\theta}{2}}{r^2} \nn\\
&&+
\fft{m\ell^4\,(5\cos2\theta-28\cos\theta-49)\, \tan^4\fft{\theta}{2}}{r^3}+
{\cal O}(r^{-4})\,.
\eea
}
Comparing with the expansions for the Bondi metric as defined in
\cite{fakenews}, we have
\bea
C_{IJ}:&& C_{\theta\theta}=0\,,\qquad C_{\phi\phi}=0\,,\qquad
C_{\theta\phi}= 4\ell\,\sin^3\fft{\theta}{2}\, \sec\fft{\theta}{2}\,,\nn\\
D_{IJ}: && D_{\theta\theta}=-4m\ell^2\,\tan^2\fft{\theta}{2}\,,\qquad
D_{\phi\phi}=16m \ell^2\,\sin^4\fft{\theta}{2}\,,\nn\\
&& D_{\theta\phi}= \fft{\ell^3}{4}\, (5\cos2\theta+12\cos\theta+15)\,
\sin^3\fft{\theta}{2}\,\sec^5\fft{\theta}{2}\,,\nn\\
C_{0\,I}:&& C_{0\, \theta}=0\,,\qquad 
C_{0\, \phi} =-4\ell \sin^2\fft{\theta}{2}\,.\label{C0CIJ}
\eea
The expressions for $C_{IJ}$ and $C_{0\,I}$ that we have obtained
here agree with the ones that follow from the procedure described in
\cite{porrati} for casting the Taub-NUT metric into a Bondi form. 
However, as far as we can judge from the higher-order terms that are
suppressed in the presentation in \cite{porrati}, their coordinate
transformation scheme will leave the metric with non-vanishing $g_{rr}$
components (starting 
at order $1/r^2$), although these should be absent in a proper
Bondi coordinate system.

Note that the tensor $C_{IJ}$ is singular at the south pole of the 2-sphere,
as is the one-form $C_{0\,I} dx^I = -4\ell\sin^2\fft{\theta}{2}\,d\phi$. 

\bibliographystyle{utphys}
\bibliography{NP}

\end{document}